# Gravitationally Induced Decoherence of Optical Entanglement

T.C.Ralph, G.J.Milburn and T.Downes

Department of Physics, University of Queensland, Brisbane 4072, QLD, Australia

(Dated: July 19, 2018)

We propose an experiment in which an entangled pair of optical pulses are propagated through non-uniform gravitational fields. A field operator calculation of this situation predicts decoherence of the optical entanglement under experimentally realistic conditions.

The effort to try to find a consistent way to combine quantum theory and gravity is made particularly difficult by a lack of experimental indicators. Typically, situations in which competing approaches make testable predictions involve experimental scenarios far beyond the reach of current technology, such as interactions with black-holes [1], worm-holes [2], highly accelerated frames [3], exotic gravitational potentials [4] or on extremely short length scales [5]. Because the non-local correlations of entangled quantum systems contrast the local nature of relativity, a fruitful area to look for experimental indicators would seem to be the behavior of entanglement in non-inertial frames [6].

Here we consider bipartite entanglement in which the members of the entangled pair experience different gravitational fields before being measured. In particular we examine the effect on optical entanglement of propagation through varying gravitational fields. Using a novel approach we predict an apparent decoherence of the entanglement under experimentally realistic conditions.

Our proposed set-up is shown in Fig.1. We consider time-energy entangled photons [7] that are sent on two different paths to a common end-point. We contrast this with classically correlated pulses sent on the same paths. The photons begin and end their journeys in regions of space with the same gravitational field. On the journey one of the photons passes through a region of significantly different gravitational field. We suppress the transverse dimensions and consider only time, $t$ and space in the longitudinal direction, $x$. We define a localized spatio-temporal mode with average frequency $\omega$ and wave number $k$, centred on the space time point $x, t$ via the mode operator:

$$\hat{a}(t,x) = \int \int dt' dx' e^{i(\omega t' - k x')} G(t' - t, x' - x) \hat{a}_{t',x'} \quad (1)$$

where $\hat{a}_{t',x'}$ are single space-time bosonic field annihilation operators with the non-zero commutators $[\hat{a}_{t',x'}, \hat{a}_{t'',x''}^\dagger] = \delta(t' - t'')\delta(x' - x'')$. $G(t', x')$ is a normalised temporal and spatial wave function, defining the measurement and interaction bandwidths of the mode. Single photon states of this mode are then given by:

$$|1\rangle_{t,x} = \hat{a}^\dagger(t,x)|0\rangle \quad (2)$$

where $|0\rangle$ is the electro-magnetic vacuum state.

The time/ position representation is related by Fourier transform to the frequency and wave-number spectra of

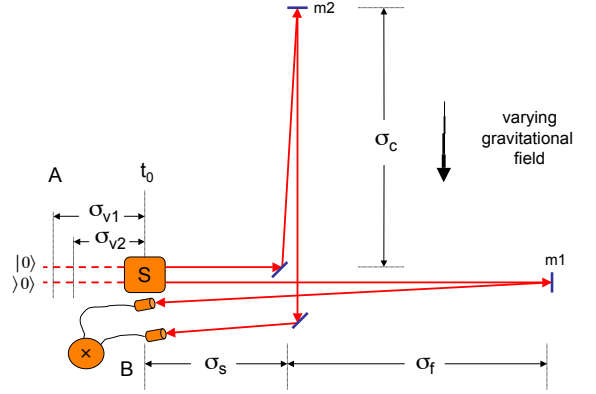

FIG. 1: Proposed experiment. Vacuum inputs at A are processed by a source S which produces two output modes. Mode 1 propagates orthogonal to the gravitational field direction till it meets mirror m1 and is reflected back to the detector at B. Mode 2 propagates anti-parallel to the field till it meets mirror m2 and it is reflected back parallel to the field to the other detector at B. The photo-currents from the detectors are combined on a multiplier to read out the coincidence current.

the mode thus an alternative representation of Eq.1 is [8]:

$$\hat{a}(t,x) = \int \int d\omega' dk' e^{i(\omega' t - k' x)} \bar{G}(\omega' - \omega, k' - k) \hat{a}_{\omega',k'}$$
$$(3)$$

where the bar indicates Fourier transform of the relevant function and the $a_{\omega',k'}$ are single wave-number/frequency bosonic annihilation operators with the non-zero commutators $[\hat{a}_{\omega',k'}, \hat{a}_{\omega'',k''}^\dagger] = \delta(\omega' - \omega'')\delta(k' - k'')$.

Free evolution over a time interval $s$ in a region of space with negligible gravitational field is generated by the unitary operator $\hat{U}_{s,l}$ which displaces the state in time by the amount $s$ and in space by $l = s$, where we work in geometric units for which the speed of light is unity. Explicitly, the unitary can be written

$$\hat{U}_{s,l} = \exp[i \int \int d\omega'' dk'' (\omega'' s - k'' l) \hat{a}_{\omega'',k''}^\dagger \hat{a}_{\omega'',k''}] \quad (4)$$

It is straightforward to confirm that $\hat{U}_{s,l}^\dagger \hat{a}(t,x) \hat{U}_{s,l} = \hat{a}(t + s, x + l)$ as expected for Heisenberg evolution of the mode operator. We wish to generalize this expression to treat radial propagation close to a massive body of mass $M$. For simplicity we will assume the body is non-spinning. The Schwarzschild metric in the radial direction for such a body is given by

$$d\tau^2 = (1 - \frac{2M}{r})dt^2 - \frac{dr^2}{(1 - \frac{2M}{r})} \qquad (5)$$

where $t$ is the time interval measured by clocks in a distant inertial frame and $r$ is the reduced circumference. However, noticing that the local proper intervals at a stationary "shell" frame at radius $r$ are given by [9]

$$\begin{aligned} ds &= \sqrt{1 - \frac{2M}{r}} dt \\ dl &= \frac{dr}{\sqrt{1 - \frac{2M}{r}}}, \end{aligned} \qquad (6)$$

we can rewrite Eq.5 as $d\tau^2 = ds^2 - dl^2$. For free optical propagation $d\tau = 0$, hence $ds = dl$ and local evolution of the mode in a shell frame is still governed by Eq.4 provided we make the substitutions $s \rightarrow ds$ and $l \rightarrow dl$. In addition, $\omega$ and $k$ in the definition of the mode operator, Eq.3, should be interpreted as the locally measured shell values of the average frequency and wave-number. The global radial evolution can be evaluated by integration over the local evolutions such that the Heisenberg evolution becomes $\hat{a}(t, x, \omega, k) \rightarrow \hat{a}(t + \int ds, x + \int dl, \omega', k')$, where $\omega' = g \, \omega$ and $k' = g^{-1} k$ with $g = \sqrt{1 - \frac{2M}{r_e}}/\sqrt{1 - \frac{2M}{r_f}}$ and $r_e$ and $r_f$ the initial and final radii respectively. In the following calculations the detectors and source will be in the same shell frame such that $r_f = r_e$ and the initial and final mode operators will have the same values of the average frequency and wave-number. We now use these tools to solve the problem posed by Fig.1.

We will refer to the shell frame in which the source and detectors are located as the SD-shell. We assume that photon mode 1 remains approximately in the SD-shell throughout its propagation, i.e. $r = r_e$ is a constant. The evolution of photon mode 2 is evaluated in a succession of shell frames as the pulse climbs out of and is then reflected back into the gravitational well. The mirrors $m1$ and $m2$ are assumed ideal reflectors across a bandwidth much larger than that of the photon modes. We set two conditions on the evolution: (i) the time intervals for the two modes to propagate to their respective mirrors and back as measured in the SD-shell are equal and; (ii) the total time between the relevant initial states and the final detection events, as determined by a sequence of shell-frame observers along the light pulse paths, are also equal. Condition (i) constrains the position of the mirrors such that classical pulses launched simultaneously from the source will strike the detectors at the same time. Condition (ii) ensures equivalent unitary evolution along all paths.

We use the Heisenberg Picture to evaluate the expected photon counting correlations by evolving the mode operators from the photon counters at $B$ back to the initial vacuum states at $A$ and expressing them as functions of the input operators. In general we will get solutions

$$\begin{aligned} \hat{a}_{m,1} &= f_1(\hat{a}_1(s, l), \hat{a}_2(s', l')) \\ \hat{a}_{m,2} &= f_2(\hat{a}_1(s'', l''), \hat{a}_2(s''', l''')) \end{aligned} \qquad (7)$$

where the functions $f_i$ describe the mixing of the mode operators and the classical pump field by the source. The different time-space labels on the operators can arise due to the different evolutions produced by the gravitational field. We assume all these times precede the source interaction, i.e. $s, s', s'' s''' << t_0$, where $t_0$ is the time at which the interaction with $S$ took place. As a result all the input operators act on the vacuum state. The rate of coincident detection events, $C$, is given by the expectation value of the product of the photon number operators, $\hat{n}_i = \hat{a}_{m,i}^\dagger a_{m,i}$ of the two modes. Thus

$$C = \langle \hat{n}_1 \hat{n}_2 \rangle = \langle 00 | \hat{a}_{m,1}^\dagger \hat{a}_{m,1} \hat{a}_{m,2}^\dagger \hat{a}_{m,2} | 00 \rangle \qquad (8)$$

In order to solve Eq.8 we must insert the correct source mode transformations and evaluate the relevant time-space labels for the mode operators given the conditions (i) and (ii). First we evaluate the integrated shell-frame time (IST), $\sigma_c$ (see Fig.1), of mode two as it climbs radially out of the field, by integrating over a succession of shell frames as it climbs. We obtain

$$\sigma_c = \int_{r_e}^{r_e + h} \frac{dr}{\sqrt{1 - \frac{2M}{r}}} \qquad (9)$$

where $r_e + h$ is the reduced circumference of mirror $m_2$. By symmetry the IST of the mode as it propagates back from the mirror to $r_e$ is also equal to $\sigma_c$. We also require the value of this same interval but as measured by clocks in the SD-shell, $\sigma_c^{(SD)}$. From Eqs 6 and the condition $ds = dl$ we have the relationship

$$dt = \frac{dr}{(1 - \frac{2M}{r})} \qquad (10)$$

Hence we obtain

$$\sigma_c^{(SD)} = \sqrt{1 - \frac{2M}{r_e}} \times \int_{r_e}^{r_e + h} \frac{dr}{(1 - \frac{2M}{r})} \qquad (11)$$

From condition (i) we require that $\sigma_f = \sigma_c^{(SD)}$ where $\sigma_f$ is the time taken by mode one in the SD-shell (see Fig.1). Condition (ii) requires that $\sigma_1 = \sigma_2$ where

$$\begin{aligned} \sigma_1 &= 2\sigma_s + 2\sigma_c + \sigma_{v1} \\ \sigma_2 &= 2\sigma_s + 2\sigma_f + \sigma_{v2} \end{aligned} \qquad (12)$$

where the various other time intervals are defined in Fig.1. Substituting from Eqs 9 and 11 we find

$$\Delta = \sigma_{v1} - \sigma_{v2} = 2(\sigma_f - \sigma_c)$$

$$
\begin{aligned}
= \ & 2(\sqrt{1 - \frac{2M}{r_e}} \times \int_{r_e}^{r_e+h} \frac{dr}{(1 - \frac{2M}{r})} \\
& - \int_{r_e}^{r_e+h} \frac{dr}{\sqrt{1 - \frac{2M}{r}}})
\end{aligned}
\tag{13}
$$

and so we can write the space time labels of mode operators evolved back via path one as

$$
s_1 = l_1 = t_0 - \sigma_{v1}
\tag{14}
$$

and that for path two as

$$
s_2 = l_2 = t_0 - \sigma_{v1} - \Delta
\tag{15}
$$

For weak fields, $2M/r << 1$, and heights much smaller than earth radius, $h/r_e << 1$, we obtain the approximate expression

$$
\Delta \approx \frac{-h^2 M}{r_e^2}
\tag{16}
$$

Now we can solve for particular sources in Fig.1. Consider first that the source S produces classically correlated coherent pulses. The displacements introduced by the classical source are assumed matched to the quantum modes. Explicitly, the input/output relationship for such a coherent source applied to mode 1 is given by

$$
\begin{aligned}
\hat{a}_{out,1} = \ & \int \int dt' dx' G(t' - t_0, x' - x_0)(\hat{a}_{1,t',x'} \\
& + G(t' - t_0, x' - x_0)\alpha) \\
= \ & \hat{a}_{in,1} + \alpha
\end{aligned}
\tag{17}
$$

where $\alpha$ is the coherent amplitude of a classical displacement field, and normalization of the spatio-temporal mode function has been used in going from the first to second line. Similarly for mode 2

$$
\hat{a}_{out,2} = \hat{a}_{in,2} + \alpha
\tag{18}
$$

where condition (i) ensures that both modes are matched to the same classical displacement. Using Eqs 14 - 18 we obtain the following expression for the evolved mode operators:

$$
\begin{aligned}
\hat{a}_{m,1} &= \hat{a}_1(s_1, l_1) + \alpha \\
\hat{a}_{m,2} &= \hat{a}_2(s_2, l_2) + \alpha
\end{aligned}
\tag{19}
$$

The mode operators appear with different time space labels, but commute for all values, and act upon their respective vacuums to produce the null state. Hence using Eq.8 we obtain

$$
C = |\alpha|^4
\tag{20}
$$

As expected the classical correlation is unaffected by propagation through the varying gravitational fields. In general this result will hold for all separable correlations.

Now we consider the source S in Fig.1 to be entangling. In particular we consider the production of time energy entanglement from the vacuum inputs via parametric down conversion. The input/output relationship for the down conversion is given by

$$
\begin{aligned}
\hat{a}_{out,1} &= \hat{a}_{in,1} + \chi \hat{a}_{in,2}^\dagger \\
\hat{a}_{out,2} &= \hat{a}_{in,2} + \chi \hat{a}_{in,1}^\dagger
\end{aligned}
\tag{21}
$$

where $\chi$ is proportional to the classical pump field of the down converter, which again is assumed mode matched to the detection modes, and we assume $\chi << 1$. In this limit the interaction described by Eqs 21 correspond to the Schrödinger evolution $|0,0\rangle \rightarrow |0,0\rangle + \chi|1\rangle_{t,x,1}|1\rangle_{t,x,2}$. This state is weakly entangled. A signature of the entanglement is that, although photon events are rare and random, they are perfectly correlated between the beams, i.e. the photons always arrive in pairs. Using Eqs 14, 15 and 21 we obtain the following expression for the evolved mode operators:

$$
\begin{aligned}
\hat{a}_{m,1} &= \hat{a}_1(s_1, l_1) + \chi \hat{a}_2^\dagger(s_1, l_1) \\
\hat{a}_{m,2} &= \hat{a}_2(s_2, l_2) + \chi \hat{a}_1^\dagger(s_2, l_2)
\end{aligned}
\tag{22}
$$

Now the commutation relations between the mode operators are changed in a non-trivial way. To second order in $\chi$ we obtain for the correlation

$$
C = |\chi[\hat{a}_1(s_1, l_1), \hat{a}_1^\dagger(s_2, l_2)]|^2
\tag{23}
$$

The commutator

$$
\begin{aligned}
& [\hat{a}_1(s_1, l_1), \hat{a}_1^\dagger(s_2, l_2)] \\
= \ & \int \int dt' dx' G(t' - s_1, x' - l_1) G(t' - s_2, x' - l_2)
\end{aligned}
\tag{24}
$$

is a real number between 0 and 1 that depends explicitly on the mode functions. However, in general it will equal 1 when $\Delta = 0$ and will equal 0 when $\Delta >> 0$. Thus in the presence of a uniform gravitational field we observe perfect photon number correlations whilst in the presence of a non-uniform gravitational field effect of sufficient magnitude those correlations completely disappear. The entanglement appears decohered.

To estimate the size of this effect we consider an experimental apparatus at sea-level where mode 2 is sent vertically a distance $h$ before being reflected back to earth. Assuming the mode function is a Gaussian of the form,

$$
G(t - t', x - x') = \sqrt{\frac{2}{\pi d_t d_x}} e^{-\frac{(t-t')^2}{d_t^2} - \frac{(x-x')^2}{d_x^2}}.
\tag{25}
$$

we obtain from Eqs 23 and 24

$$
C = \chi^2 e^{-(\frac{\Delta^2}{d_t^2} + \frac{\Delta^2}{d_x^2})}.
\tag{26}
$$

We estimate the *intrinsic* temporal and spatial uncertainties of a silicon photon counter to be 300 fs and 1

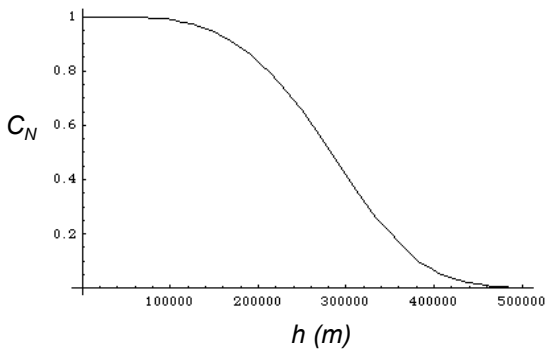

$C_N$

$h$ (m)

mm resectively. Hence, we set the standard deviations in Eq.25 to $d_t = 10^{-5} m$ and $d_x = 10^{-3} m$. These numbers also roughly correspond to the pulse duration and pump spot radius for typical, single pass, pulsed parametric down-conversion. The results of these numerical calculations are shown in Fig.2. We find strong decoherence (i.e. disappearance of coincidences) for $h > 400$ kilometers.

The novelty of the predicted effect should not be underestimated. First note that although, because of the loss of photon correlations, we refer to this effect as decoherence, in fact the effect is in principle reversible by resending (before detection) mode 1 along mode 2's path and vice versa. Secondly we note that energy conservation, as usually expected from down conversion, is only satisfied on average, not shot by shot. On the other hand in the current model both the down conversion pump and the gravitational field itself are treated as un-depleted classical energy reservoirs, so a definitive statement on this issue is beyond the scope of the present model. Thirdly we anticipate that more unusual evolutions may arise for strongly entangled qubit states as suggested in Ref [10]. Treatment of such situations with the same rigour as used here would require consideration of highly nonlinear Heisenberg evolutions that are, again, beyond the scope of the present calculations.

We have studied the effect on optical entanglement of evolution through varying gravitational fields. We have predicted a decoherence effect that should be observable under experimentally achievable conditions. We believe the outcome of such an experimental investigation could have considerable implications for the unification of quantum mechanics and general relativity. Furthermore, the predicted effect, if observed, would represent a new phenomenon with major consequences for quantum physics in general and quantum information in particular.

We wish to thank David Pulford, Paul Davies, Daniel Gottesman and Craig Savage for useful discussions. This work was supported by the Australian Research Council.